\documentclass[prd,eqsecnum,twocolumn,amsfonts,amssymb]{revtex4}

\usepackage{graphicx}

\usepackage{bm}

\newcommand{\beq}{\begin{equation}}
\newcommand{\eeq}{\end{equation}}
\newcommand{\beqs}{\begin{eqnarray}}
\newcommand{\eeqs}{\end{eqnarray}}

\newcommand{\Deltaslash}{\Delta\hspace{-0.065in}\slash}

\begin{document}

\title{Scheme-Independent Calculations of Anomalous Dimensions of Baryon
  Operators in Conformal Field Theories}

\author{John A. Gracey$^a$, Thomas A. Ryttov$^b$ and Robert Shrock$^c$}

\affiliation{(a) \ Theoretical Physics Division, Department of Mathematical
  Sciences, \\ University of Liverpool, \\ P.O. Box 147, Liverpool L69 3BX,
  United Kingdom}

\affiliation{(b) \ CP$^3$-Origins, University of Southern Denmark,
 Campusvej 55, DK-5230 Odense M, Denmark}

\affiliation{(c) \ C. N. Yang Institute for Theoretical Physics and
Department of Physics and Astronomy, \\
Stony Brook University, Stony Brook, NY 11794, USA }

\begin{abstract}

  We present the first analytic scheme-independent series calculations of
  anomalous dimensions of several types of baryon operators at an infrared
  fixed point (IRFP) in an asymptotically free SU(3) gauge theory with $N_f$
  fermions.  Separately, for an asymptotically free gauge theory with a gauge
  group $G$ and $N_f$ fermions in a representation $R$ of $G$, we consider
  physical quantities at an IRFP, including the anomalous dimension of
  gauge-invariant fermion bilinears and the derivative of the beta
  function. These quantities have been calculated in series expansions whose
  coefficients have been proved to be scheme-independent at each order. We
  illustrate the scheme independence using a variety of schemes, including the
  RI$^\prime$ scheme and several types of momentum subtraction (MOM) schemes.

\end{abstract}

\maketitle


\section{Introduction}
\label{intro_section}

In conformal field theories, quantitites of particular interest are the scaling
dimensions, $D_{\cal O}$, of gauge-invariant operators, ${\cal O}$. In general,
we write
\beq
D_{\cal O} = D_{{\cal O},cl.} - \gamma_{\cal O} \ ,
\label{dimop}
\eeq
where $D_{{\cal O},cl.}$ is the classical (free-field) dimension of ${\cal O}$,
and $\gamma_{\cal O}$ is the anomalous dimension of ${\cal O}$ due to
interactions.  We shall focus on the determination of $\gamma_{\cal O}$ in
perturbation theory at a fixed point of the renormalization group (RG). An
important example of such a fixed point is encountered in the case of an
asymptotically free non-Abelian gauge theory with gauge group $G$ and
sufficiently many massless fermions in a representation $R$ of $G$.  We denote
the running gauge coupling at a Euclidean scale $\mu$ as $g=g(\mu)$ and denote
$\alpha=g^2/(4\pi)$ and $a=g^2/(16\pi^2)$. In this theory, the gauge coupling
evolves from small values in the ultraviolet (UV) at large $\mu$ to an infrared
fixed point (IRFP) at a value denoted $\alpha_{IR}$ as $\mu \to 0$. At this
value, the theory is scale-invariant and is inferred to be conformally
invariant \cite{scalecon,fm}. This infrared behavior is commonly denoted the
non-Abelian Coulomb phase (NACP) or conformal window. The RG evolution of the
gauge coupling is described by the beta function,
\beq
\beta_\alpha = \frac{d\alpha}{d\ln\mu} = -2\alpha 
\sum_{\ell=1}^\infty b_\ell \, a^\ell \ , 
\label{beta} 
\eeq
where $b_\ell$ is the $\ell$-loop coefficient.  At the two-loop $(2\ell)$ level
\cite{b1}-\cite{bz}, the IR zero of $\beta_\alpha$ function occurs at
\beq
\alpha_{IR,2\ell}=-\frac{4\pi b_1}{b_2} \ .
\label{alfir_2loop}
\eeq
If $N_f$ is only slightly smaller than the upper limit, 
\beq
N_u=\frac{11C_A}{4T_f}
\label{nfb1z}
\eeq
implied by the property of asymptotic freedom \cite{casimir,nfintegral}, then
$\alpha_{IR,2\ell}$ is small and can be analyzed perturbatively \cite{b2,bz}.
As $N_f$ decreases, the value of the coupling at the infrared zero of the beta
function increases, motivating calculation of this IRFP value of $\alpha$ to
higher-loop order. This was carried out to the four-loop level for general
gauge group $G$ and fermion representation $R$ in \cite{bvh}-\cite{bc}, using
$b_3$ \cite{b3} and $b_4$ \cite{b4} computed in the modified minimal
subtraction scheme \cite{msbar} scheme for regularization and renormalization,
denoted $\overline{\rm MS}$.  (The minimal subtraction scheme was originally
presented in \cite{ms}.)  Subsequently, the IRFP was calculated to the
five-loop level \cite{flir}, using $b_5$ in the $\overline{\rm MS}$ scheme
\cite{b5su3,b5}.  Effects of scheme dependence were studied in
\cite{sch}-\cite{bzmom2015}.

The anomalous dimension of a gauge-invariant operator ${\cal O}$ is, in
principle, measurable, and hence cannot depend on the scheme used for
regularization and renormalization.  However, this property is not maintained
in a conventional finite-order perturbative calculation of the anomalous
dimension of such an operator as a power series in the coupling $\alpha$,
\beq
\gamma_{\cal O} = \sum_{\ell=1}^\infty c_{{\cal O},\ell} \, a^\ell \ . 
\label{gammao}
\eeq
Once the perturbative expansion for $\gamma_{\cal O}$ is truncated at a finite
order, scheme dependence is induced in the result for $\gamma_{\cal O}$. Only
if one had the entire perturbative series available would the final result be
guaranteed to be scheme-independent. Explicitly, to evaluate $\gamma_{\cal O}$
to finite order at an IRFP using Eqs. (\ref{beta}) and (\ref{gammao}), one
solves for the relevant zero of the $n$-loop beta function to obtain the
$n$-loop value of $\alpha$ at this IRFP, denoted $\alpha_{IR,n\ell}$ and then
substitutes this into Eq. (\ref{gammao}) to obtain the value at the IRFP,
$\gamma_{{\cal O},IR}$.  However, beyond the lowest orders, the result is
scheme-dependent, because of scheme dependence in both the higher-order 
$b_\ell$ and the $c_{{\cal O},\ell}$ coefficients.  The calculations of
$\gamma_{\bar\psi\psi,IR}$ to four-loop order in \cite{bvh,ps} and to five-loop
order in \cite{flir} used the four-loop and five-loop coefficients
$c_{\bar\psi\psi,4}$ \cite{c4} and $c_{\bar\psi\psi,5}$ \cite{c5},
respectively, calculated in the $\overline{\rm MS}$ scheme. This scheme
dependence of higher-order perturbative calculations is, of course, not limited
to these quantities, but is a generic property of higher-order
calculations. For example, it is well known that higher-order calculations of
differential and total cross sections in quantum chromodynamics (QCD) are also
scheme-dependent.

Intuitively, one expects that as one increases the order of the perturbative
computation, there is more scheme-independent information contained in
$\gamma_{\cal O}$. This expectation is justified by the fact that higher-order
QCD calculations used, e.g., to analyze data from the Fermilab Tevatron and
CERN Large Hadron Collider showed less dependence on the scheme/scale than
lower-order calculations \cite{qcdreviews}. Indeed, for many years there has
been work on the construction and application of schemes in QCD designed to
reduce the scheme and scale dependence in higher-order QCD calculations (e.g.,
\cite{celmaster}-\cite{brodsky}).

Ideally, one would use a method of perturbative calculation of physical
quantities that manifestly preserves the scheme independence at each finite
order in the series expansion. That is, one would like to extract the
scheme-independent information that is contained in the scheme-dependent
higher-order coefficients $b_\ell$ and $c_{{\cal O},\ell}$.  A key property of
the IRFP in an asymptotically free gauge theory is that $\alpha_{IR} \to 0$ as
$N_f$ (considered to be generalized to real numbers \cite{nfintegral})
approaches the upper limit, $N_u$, allowed by asymptotic freedom.  It follows
that one can reexpress a physical quantity such as $\gamma_{\cal O}$ at the
IRFP as a series expansion in powers of the difference
\beq
\Delta_f = N_u - N_f \ , 
\label{delta}
\eeq
i.e., 
\beq
\gamma_{\cal O} = \sum_{n=1}^{\infty}  \kappa_{{\cal O},n} \Delta_f^n \ . 
\label{gamma_dex}
\eeq
Since $\Delta_f$ is obviously scheme-independent and so is $\gamma_{\cal O}$,
each coefficient $\kappa_{{\cal O},n}$ is also scheme-independent. Some early
work based on this was in \cite{bz,gkgg}.

Recently, extensive scheme-independent expansions for anomalous dimensions of a
number of physical quantities have been calculated and analyzed in
\cite{gtr}-\cite{dualsqcd}.  For asymptotically free vectorial gauge theories
with gauge group $G$ and $N_f$ fermions transforming according to a
representation $R$ of $G$, physical quantities of interest include 
the fermion bilinears $\bar\psi\psi$ and
$\bar\psi {\cal T}_j\psi$, where we suppress the sum over fermion flavor
indices and ${\cal T}_j$ denotes a generator of the Lie algebra of
SU($N_f$). These have the same anomalous dimension
\cite{gracey_gammatensor}. We denote this anomalous dimension as
$\gamma_{\bar\psi\psi}$ and its evaluation at the IRFP as
$\gamma_{\bar\psi\psi,IR}$.  The scheme-independent series expansion of 
$\gamma_{\bar\psi\psi,IR}$ is written as 
\beq
\gamma_{\bar\psi\psi,IR} = \sum_{n=1}^\infty \kappa_n \, \Delta_f^n \ . 
\label{gamma_deltaseries}
\eeq
In general, the calculation of the coefficient $\kappa_n$ in
Eq. (\ref{gamma_deltaseries}) requires, as inputs, the values of the $b_\ell$
for $1 \le \ell \le n+1$ and the $c_\ell$ for $1 \le \ell \le n$.

The derivative of the beta function evaluated at the IRFP, 
\beq
\beta'_{IR} = \frac{d\beta_\alpha}{d\alpha} \bigg |_{\alpha=\alpha_{IR}}
\label{betaprime}
\eeq
is also a physical quantity and hence is scheme-independent \cite{gross75}. 
Indeed, from the trace anomaly \cite{traceanomaly} $T^\mu_\mu =
[\beta_\alpha/(4 \alpha)] F_{\mu\nu}^a F^{a \mu\nu}$, where $F_{\mu\nu}^a$ is
the field-strength tensor, it follows that the full scaling dimension of
$F^2 \equiv {\rm Tr}(F_{\mu\nu}F^{\mu\nu})$, satisfies the relation
\cite{gubser}
\beq
D_{F^2}=4+\frac{d\beta_\alpha}{d\alpha}-\frac{2}{\alpha}\, \beta_\alpha \ ,
\label{dfsq}
\eeq
so that, at the IRFP, with $\beta=0$, $\gamma_{F^2,IR}=-\beta'_{IR}$, i.e.,
$\beta'_{IR}$ is equivalent to the anomalous dimension of $F^2$ evaluated at
the IRFP.  The scheme-independent series expansion of $\beta'_{IR}$ is written
as 
\beq
\beta'_{IR} = \sum_{n=2}^\infty d_n \, \Delta_f^n 
\label{betaprime_deltaseries}
\eeq
(Note that $d_1=0$ for all $G$ and $R$.)
In general, the calculation of the coefficient $d_j$
in Eq. (\ref{betaprime_deltaseries}) requires, as inputs, the values of the
$b_\ell$ for $1 \le \ell \le j$. 
In addition to these calculations for vectorial gauge theories,
Ref. \cite{bpcgt} carried out scheme-independent calculations of $\beta'_{IR}$
for chiral gauge theories.

The results of the scheme-independent series expansions in
\cite{gtr}-\cite{pgb} are useful for several reasons, which are also
motivations for the present study.  First, they give new information about
fundamental properties of conformal field theories, namely anomalous dimensions
at an IRFP in the non-Abelian Coulomb phase of an asymptotically free gauge
theory.  A second important use of these calculations pertains to the
determination of the size of the NACP.  The upper end of the NACP, as a
function of $N_f$, is known and is equal to $N_u$.  However, for
non-supersymmetric theories, the lower end, at a value that we denote as
$N_{f,cr}$, is not known, and there is an intensive ongoing effort to determine
$N_{f,cr}$ by means of lattice simulations \cite{lgtreviews,simons}. Applying
scheme-independent calculations of $\gamma_{\bar\psi\psi,IR}$,
Refs. \cite{gsi,dex,dexs,dexl,dexo,pgb} obtained estimates of $N_{f,cr}$ in a
manner complementary to lattice gauge simulations. This was done using the
monotonic increase of $\gamma_{\bar\psi\psi,IR}$ with decreasing $N_f$ that was
shown by the scheme-independent calculations, in conjunction with the rigorous
upper limit on $\gamma_{\bar\psi\psi,IR}$ from conformal invariance, namely
$\gamma_{\bar\psi\psi,IR} < 2$ \cite{gammabound}. A third application follows
from the second, namely that a knowledge of $N_{f,cr}$ (for a given gauge group
$G$ and fermion representation $R$) is necessary for the construction and study
of quasiconformal theories of physics beyond the Standard Model (BSM), since
these require $N_f$ to be slightly less than $N_{f,cr}$ in order to achieve the
slow running of the gauge coupling and associated quasiconformal behavior.  In
turn, the dynamical breaking of the approximate dilatation invariance in these
theories leads to a light approximate Nambu-Goldstone boson, the dilaton
\cite{dil,ggs,dillat,lgtreviews,simons}. These vectorial BSM theories can
naturally arise from the sequential breaking of asymptotically free chiral
gauge theories \cite{etc}.  This is relevant to the investigation of the Higgs
boson; although its production and decay properties are consistent with the
predictions of the Standard Model, there is the continuing question of whether
it might be a composite, dilaton-like state resulting from a quasiconformal BSM
theory \cite{ggs}.

The accuracy of the scheme-independent series expansions of
$\gamma_{\bar\psi\psi,IR}$ and $\beta'_{IR}$ was studied in several ways in
\cite{gtr}-\cite{dualsqcd}.  One way 
was to evaluate the stability of these quantities as higher-order terms in
powers of $\Delta_f$ were added in the series.  It was shown that the
finite-order scheme-independent series calculations were most accurate at the
upper end of the NACP, and remained reasonably accurate over a substantial
portion of the NACP extending to lower values of $N_f$.

For the gauge group $G={\rm SU}(3)$, a baryon operator has the form of a
product of three fermion fields, each transforming as the fundamental (triplet)
representation of $G$, with their gauge indices $a,b,c$ contracted with the
$\epsilon_{abc}$ tensor to form a color singlet.  Relevant previous studies of 
anomalous dimensions of baryon operators in QCD include
\cite{peskin}-\cite{vecchi}.  In particular, the anomalous dimensions of
baryon operators have been calculated to one-loop \cite{peskin}, two-loop
\cite{pivo,krankl_manashov}, and three-loop order \cite{gracey_baryon,gbe} as
powers series in $\alpha$ and related studies have been presented in 
\cite{psbaryon,vecchi}. 

In this paper we shall present, for the first time, analytic scheme-independent
series calculations to order $O(\Delta_f^3)$ of anomalous dimensions of several
types of baryon operators at an infrared fixed point of an asymptotically free
SU(3) gauge theory with $N_f$ fermions in the fundamental representation. An
assessment of the accuracy of these calculations will also be given. As was
discussed previously \cite{gtr,gsi,dex}, the procedure for the calculation of
scheme-independent series expansions requires that the IRFP be exact, and this
is only the case in the non-Abelian Coulomb phase, in which the chiral flavor
symmetry is exact \cite{sksb}. Since we thus necessarily restrict our analysis
to the NACP, where there is no confinement, we use the term ``baryon'' to refer
only to the property that the baryon operators that we consider are singlets
under the SU(3) gauge symmetry.  We note that there is actually some irony in
using the term ``baryon'' here, since it is derived from the Greek word $\beta
\alpha \rho \upsilon \varsigma$, meaning ``heavy''. However, a gauge-singlet
state produced by the operation of a baryon creation operator on the vacuum in
the non-Abelian Coulomb phase is massless, as are all physical states in this
phase.

As a second part of our paper, we shall present, 
for general gauge group $G$ and fermion representation $R$, an
explicit illustration of the scheme independence of the earlier calculations of
$\Delta_f$ expansions of $\gamma_{\bar\psi\psi,IR}$ and $\beta'_{IR}$
\cite{gtr,gsi,dex,dexs,dexl,dexo,pgb}.  These calculations naturally used the
$\overline{\rm MS}$ scheme because the $n$-loop coefficients in the beta
function and in $\gamma_{\bar\psi\psi}$ had been calculated to the highest loop
order in this scheme, and these coefficients have the simplest form in this
scheme. Since a rigorous proof was already given in these earlier works of the
scheme independence of the coefficients in these $\Delta_f$ expansions, it is
not necessary to carry out the calculations in schemes other than the simplest
one.  However, it is, nevertheless, quite instructive to see how the
considerably more complicated higher-order coefficients in the beta function
and anomalous dimensions in these more complicated schemes combine to reproduce
exactly the results of the $\overline{\rm MS}$ scheme for the coefficients in
the various $\Delta_f$ series expansions.  For the purpose of these
illustrations, we shall consider a variety of different schemes, including the
RI$^\prime$ scheme \cite{ri_prime,gracey_ri_prime} and several varieties of
momentum (MOM) subtraction schemes
\cite{celmaster},\cite{mmom}-\cite{graceysimms_mom_omega} (see also \cite{r5}).

It should be mentioned that this program of explicitly demonstrating scheme
independence of the coefficients in the $\Delta_f$ expansions of 
anomalous dimensions of various operators 
was previously carried out for the ${\cal N}=1$ supersymmetric gauge
theories in \cite{gtr,dex,dexs,dexss}, where it was shown that the use of two
different schemes, namely the $\overline{\rm DR}$ scheme \cite{drbar} and the
Novikov-Shifman-Vainshtein-Zakharov (NSVZ) scheme 
\cite{nsvz} yield the same scheme-independent results for the anomalous
dimension of a holomorphic composite product of chiral superfields,
$\gamma_{\Phi_{comp.},IR}$, which, order-by-order are in precise agreement with
the corresponding series expansion of the exactly known expression
\cite{seiberg}. In addition to demonstrating explicitly that different schemes
yield the same values of coefficients in the scheme-independent expansion of
$\gamma_{\Phi_{comp.},IR}$ of the form (\ref{gamma_dex}), this work showed that
(i) the series (\ref{gamma_dex}) converges to the exact expression everywhere
where the latter applies, i.e., in the NACP, (ii) for a fixed $N_f$ in the
NACP, a finite truncation of the series (\ref{gamma_dex}) to order
$O(\Delta_f^p)$ approaches the exact expression exponentially rapidly, and
(iii) throughout the entire NACP, one achieves excellent accuracy of a few
percent even with a series calculated to a modest order of $n=4$, i.e.,
$O(\Delta_f^4)$. These scheme-independent calculations of anomalous dimensions
in an ${\cal N}=1$ supersymmetric gauge theory thus improved upon
conventional scheme-dependent series expansions in powers of $\alpha_{IR}$
\cite{bfs}-\cite{bfss} (see also \cite{kataev}). 


\section{Baryon Operators} 
\label{baryon_operators_section}

In this section we consider a theory with gauge group $G={\rm SU}(3)$ and $N_f$
fermions in the fundamental (triplet) representation, $R=F$.  Since the
fermions are massless, the ultraviolet theory is invariant under the global
flavor ($fl.$) symmetry group
\beq
G_{fl.} = {\rm SU}(N_f)_L \otimes {\rm SU}(N_f)_R \otimes {\rm U}(1)_V \ .
\label{gfl}
\eeq
This symmetry is unbroken in the non-Abelian Coulomb phase. Hence, the baryon
operators that we consider transform according to definite representations of
this group.  Each fermion field can be decomposed into its left- and
right-handed chiral components as $\psi = (P_L+P_R)\psi = \psi_L + \psi_R$,
where $P_{R,L}=(1/2)(1 \pm \gamma_5)$ and we suppress color and flavor indices
here.  Showing these latter indices explicitly, each fermion field can thus be
written formally as $\psi^a_{i,L} + \psi^a_{i,R}$, where $a$ is an SU(3) color
gauge index. Here, the flavor index $i$ on $\psi^a_{i,L}$ refers to the
fundamental representation of ${\rm SU}(N_f)_L$, while the flavor index $i$ 
on $\psi^a_{i,R}$ refers to the fundamental representation of 
${\rm SU}(N_f)_R$. This will be understood implicitly below. 
The chiral components $\psi^a_{i,L}$ and $\psi^a_{i,R}$
transform as $(N_f,1)$ and $(1,N_f)$ under the chiral part of $G_{fl.}$, ${\rm
  SU}(N_f)_L \otimes {\rm SU}(N_f)_R$. The bilinear operator
$\bar \psi \psi = \sum_{i=1}^{N_f}(\bar\psi_{i,L} \psi_{i,R} + 
\bar\psi_{i,R} \psi_{i,L})$ thus corresponds to what would be the 
flavor-singlet in the confined phase, where the chiral part of 
$G_{fl.}$ is broken to the diagonal ${\rm SU}(N_f)_V$ subgroup, while the
operator $\bar \psi {\cal T}_j \psi$ corresponds to what would be the 
flavor-adjoint in the confined phase. In our present work 
we will use the symbols $S_{k,L}$ and $A_{k,L}$ to denote the 
$k$-fold symmetric and $k$-fold antisymmetric representations of 
${\rm SU}(N_f)_L$, and similarly with  $S_{k,R}$ and $A_{k,R}$ 
with ${\rm SU}(N_f)_R$.  

Clearly, all of our baryon operators have unit baryonic charge under
the U(1)$_V$ factor group (which is equivalent to U(1)$_B$ here) so we leave
this implicit henceforth.  Although we are in an NACP without any confinement
of color, it is nonetheless convenient to deal with gauge-singlet operators,
since they are gauge-invariant.  The invariance of the baryon operator under
the SU(3) gauge group is guaranteed by the contraction of the color indices
$a,b,c$ on the three fermion fields with the $\epsilon_{abc}$ tensor, so that
the color part of the baryon wavefunction is totally antisymmetric.  The other
parts of the baryon operator depend on the chirality, spin contractions, and
flavor structure of the three-fermion operator. These are constrained by the
requirement that the full wavefunction must be totally antisymmetric under
interchange of any two of the fermions.

As is well known, relevant representations of the Lorentz group SO(3,1) are
specified by two spins, $(j_1,j_2)$. It is convenient to construct a subset of
baryon operators by combining two of the three fermions in a Majorana-type
bilinear operator product, since this has spin 0 and is Lorentz-invariant. A
Majorana-type bilinear links left-handed to left-handed chiral components of a
fermion, and right-handed to right-handed chiral components. There are thus two
of these, namely $\psi^{a \ T}_{i,L} C \psi^b_{j,L}$ and $\psi^{a \ T}_{i,R}
\psi^b_{j,R}$. Here, $C$ is the Dirac charge conjugation matrix defined by $C
\gamma_\mu C^{-1} = -\gamma_\nu^T$ and satisfying the properties $C^T=-C$ and
$C^{-1}=C^T$.  The full baryon operator product is then obtained by combining
each of these Majorana-type bilinears with the left-handed or right-handed
chiral fermion.  One thus has the operators
\beq
{\cal O}_{RLL} = 
\epsilon_{abc}\psi^a_{i,R} [\psi_{j,L}^{b \ T} C \psi^c_{k,L}]
\label{o_rll}
\eeq
\beq
{\cal O}_{LRR} = 
\epsilon_{abc}\psi^a_{i,L} [\psi_{j,R}^{b \ T} C \psi^c_{k,R}]
\label{o_lrr}
\eeq
\beq
{\cal O}_{RRR} = 
\epsilon_{abc}\psi^a_{i,R} [\psi_{j,R}^{b \ T} C \psi^c_{k,R}]
\label{o_rrr}
\eeq
and
\beq
{\cal O}_{LLL} = 
\epsilon_{abc}\psi^a_{i,L} [\psi_{j,L}^{b \ T} C \psi^c_{k,L}] \ . 
\label{o_lll}
\eeq
To distinguish the chirality of the unpaired fermion, one could use
a subscript $L$ or $R$, but we shall follow the notational conventions of
\cite{pivo,gracey_baryon}, according to which
\beq
{\cal O}^{(\frac{1}{2},0)}_+ \equiv {\cal O}^{(\frac{1}{2},0)}_{LLL}
\label{o_120p}
\eeq
and
\beq
{\cal O}^{(\frac{1}{2},0)}_- \equiv {\cal O}^{(\frac{1}{2},0)}_{RLL} \ . 
\label{o_120m}
\eeq
As is evident, in the Lorentz $(j_1,j_2)$ labelling, the $j_1=1/2$ refers to
the fermion field that is not a member of the Majorana fermion bilinear, and
$j_2=0$ refers to the spin-0 transformation property of this Majorana fermion
bilinear. These operators have anomalous dimensions denoted 
$\gamma_B^{(\frac{1}{2},0),+}$ and $\gamma_B^{(\frac{1}{2},0),-}$,
respectively. Because the theory at the IRFP in the non-Abelian phase 
preserves the full flavor symmetry (\ref{gfl}), the anomalous dimension 
$\gamma_B^{(\frac{1}{2},0),+}$ for ${\cal O}^{(\frac{1}{2},0)}_{LLL}$ is equal
to the anomalous dimension for the corresponding operator with all $L$ indices
switched to $R$, namely ${\cal O}^{(\frac{1}{2},0)}_{RRR}$, and, separately,
the anomalous dimension $\gamma_B^{(\frac{1}{2},0),-}$ for 
${\cal O}^{(\frac{1}{2},0)}_{RLL}$ is equal
to the anomalous dimension for the corresponding operator with $L$ and $R$ 
indices interchanged, namely ${\cal O}^{(\frac{1}{2},0)}_{LRR}$.

One part of the classification of baryon operators entails the analysis of the
combination of the three spin 1/2 representations of angular momentum SU(2). In
general, one has
\beq
\frac{1}{2} \times \frac{1}{2} \times \frac{1}{2} = 
\frac{1}{2} + \frac{1}{2} + \frac{3}{2}
\label{sss} 
\eeq
(i.e., $2 \times 2 \times 2 = 2 + 2 + 4$ in terms of the dimensions $2s+1$ of
the representations). We have considered above the cases in which two of the
spins are contracted to produce spin 0, corresponding to one of the two
spin-1/2 terms on the right-hand side of Eq. (\ref{sss}).  There are two
remaining cases to consider, in which one combines two of the spins to produce
a spin-1 state and then combines this with the third spin 1/2 to yield a net
spin 1/2 or spin 3/2.  We recall that the spin wavefunction in the case of spin
3/2 is totally symmetric, i.e., $S_3$ under the SU(2) of spin.  In the analysis
of baryon operators in QCD, it has proved useful to introduce a vector
$\Delta_\mu$ that is lightlike, i.e., has the property $\Delta^2=0$, and
consider the operators (leaving the flavor indices implicit) 
\beq
{\cal O}^{(\frac{3}{2},0)}_{LLL} =
\epsilon_{abc}\Deltaslash \psi^a_{L} 
              \Deltaslash \psi^b_{L}
              \Deltaslash \psi^c_{L}
\label{odelta_320_lll}
\eeq
\beq
{\cal O}^{(\frac{3}{2},0)}_{RRR} = 
\epsilon_{abc}\Deltaslash \psi^a_{R} 
              \Deltaslash \psi^b_{R}
              \Deltaslash \psi^c_{R}
\label{odelta_320_rrr}
\eeq
\beq
{\cal O}^{(1,\frac{1}{2})}_{LLR} =
\epsilon_{abc}\Deltaslash \psi^a_{L} 
              \Deltaslash \psi^b_{L}
              \Deltaslash \psi^c_{R}
\label{odelta_112_llr}
\eeq
and
\beq
{\cal O}^{(1,\frac{1}{2})}_{RRL} = 
\epsilon_{abc}\Deltaslash \psi^a_{R} 
              \Deltaslash \psi^b_{R}
              \Deltaslash \psi^c_{L} \ . 
\label{odelta_112_rrl}
\eeq
In the notation of \cite{pivo,gracey_baryon}, 
\beq
{\cal O}^{(\frac{3}{2},0)}_+ \equiv {\cal O}^{(\frac{3}{2},0)}_{LLL}
\label{odelta_320p}
\eeq
and
\beq
{\cal O}^{(1,\frac{1}{2})}_- \equiv {\cal O}^{(1,\frac{1}{2})}_{LLR} \ . 
\label{odelta_112m}
\eeq

The anomalous dimensions of these operators are denoted 
$\gamma^{(\frac{3}{2},0),+}$ and $\gamma^{(1,\frac{1}{2}),-}$, respectively. 
Again, owing to the exact chiral symmetry (\ref{gfl}), the anomalous dimension 
$\gamma^{(\frac{3}{2},0),+}$ of ${\cal O}^{(\frac{3}{2},0)}_{LLL}$ is equal to
the anomalous dimension of ${\cal O}^{(\frac{3}{2},0)}_{RRR}$, and the
anomalous dimension $\gamma^{(1,\frac{1}{2}),-}$ of 
${\cal O}^{(1,\frac{1}{2})}_{LLR}$ is equal to
the anomalous dimension of ${\cal O}^{(1,\frac{1}{2})}_{RRL}$. The
normalization of these anomalous dimensions is fixed by the basic relation
(\ref{dimop}). 


\section{Scheme-Independent Series Expansion for Anomalous Dimension of 
General Baryon Operator}
\label{gamma_baryon_section}

A general expression, calculated to the two-loop level, was given for the
anomalous dimension of a general baryon operator ${\cal O}_B$ in
\cite{krankl_manashov} and extended to the three-loop level in
\cite{gracey_baryon,gbe}. This depends on certain coefficients ${\mathbb C}_k$,
which are listed in Table \ref{cvalues}.  With the definition (\ref{dimop})
(which sets the absolute normalization of the anomalous dimension), and noting
that the sign convention in (\ref{dimop}) is opposite to that in
\cite{gracey_baryon}, we have 
\begin{widetext}
\beqs 
\gamma_{B} &=& \frac{1}{3}{\mathbb C}_2 \, a 
+ \left [ (-72+4N_f){\mathbb C}_0
+\left ( \frac{47}{18}-\frac{1}{27}N_f \right ){\mathbb C}_2 
+ \frac{1}{36}{\mathbb C}_2^2 - \frac{5}{36}{\mathbb C}_4 \right ] \, a^2
\cr\cr
&+& \Bigg [ \left ( -\frac{16094}{9} - 34\zeta_3 + \frac{1706}{9}N_f
    -\frac{20}{9}N_f^2 \right ){\mathbb C}_0
+ \left ( \frac{5873}{108} - \frac{433}{18}\zeta_3 -\left ( \frac{71}{27} +
\frac{40}{9}\zeta_3 \right )N_f - \frac{13}{81}N_f^2 \right ) {\mathbb C}_2
\cr\cr
&+& \left (-\frac{209}{324}+\frac{71}{27}\zeta_3+\frac{1}{324}N_f \right )
{\mathbb C}_2^2
+ \left ( \frac{5}{648} - \frac{1}{27}\zeta_3 \right ){\mathbb C}_2^3 \cr\cr
&+& \left ( \frac{91}{72} - \frac{29}{12}\zeta_3 + \frac{7}{324}N_f \right )
{\mathbb C}_4
+ \left ( -\frac{37}{432} + \frac{25}{144}\zeta_3 \right )
{\mathbb C}_2{\mathbb C}_4
+ \left ( -\frac{1}{8}+ \frac{2}{9}\zeta_3 \right ) {\mathbb C}_{444} 
\Bigg ] \, a^3 + O(a^4) \ . 
\label{gamma_generalbaryon}
\eeqs
\end{widetext}
We list the values of the ${\mathbb C}_k$ coefficients for various specific
baryon operators in Table \ref{cvalues}.  

\begin{table}
\caption{\footnotesize{Values of ${\mathbb C}_k$ coefficients.}}
\begin{center}
\begin{tabular}{|c|c|c|c|c|c|} \hline\hline
$(j_1,j_2)$ & chirality & ${\mathbb C}_0$ & ${\mathbb C}_2$
& ${\mathbb C}_4$ & ${\mathbb C}_{444}$
\\ \hline
$(\frac{1}{2},0)$ & $+$ & 1  & 12    &    72   & 0   \\
$(\frac{1}{2},0)$ & $-$ & 1  & 12    &  $-24$  & 0   \\
$(\frac{3}{2},0)$ & $+$ & 1  & $-12$ &    72   & 0   \\
$(1,\frac{1}{2})$ & $-$ & 1  & $-4$  &  $-24$  & 0   \\
\hline\hline
\end{tabular}
\end{center}
\label{cvalues}
\end{table}

We denote the anomalous dimension of the general baryon operator ${\cal O}_B$
as $\gamma_{{\cal O}_B}$ and write the scheme-independent series expansion for
this as 
\beq
\gamma_B = \sum_{n=1}^\infty \kappa_{B,n} \, \Delta_f^n \ . 
\label{gamma_baryon_deltaseries}
\eeq
For this SU(3) gauge theory with $N_f$ fermions in the fundamental
representation, $N_u = 33/2$, so the general expression for $\Delta_f$ in
Eq. (\ref{delta}) yields $\Delta_f=(33/2)-N_f$. 

We calculate the following coefficients in this scheme-independent series
expansion for the general baryon operator:
\beq
\kappa_{B,1} = \frac{2}{3^2 \cdot (107)} {\mathbb C}_2 \ , 
\label{kappa1_generalbaryon}
\eeq
\begin{widetext}
\beq
\kappa_{B,2} = -\frac{8}{3 \cdot (107)^2}{\mathbb C}_0 
+ \frac{27083}{2 \cdot 3^4 \cdot (107)^3}{\mathbb C}_2 
+ \frac{1}{3^4 \cdot (107)^2} ({\mathbb C}_2^2-5{\mathbb C}_4) \ , 
\label{kappa2_generalbaryon}
\eeq
and
\beqs
\kappa_{B,3} &=& \left ( \frac{291892}{3^5 \cdot (107)^4}
- \frac{272}{3^3 \cdot (107)^3}\zeta_3 \right ){\mathbb C}_0  
+ \left ( \frac{352124197}{2^2 \cdot 3^6 \cdot (107)^5} - 
\frac{238124}{3^5 \cdot (107)^4}\zeta_3 \right ) {\mathbb C}_2 \cr\cr
&+& \left ( -\frac{47365}{2 \cdot 3^7 \cdot (107)^4} + 
\frac{568}{3^6 \cdot (107)^3}\zeta_3 \right ){\mathbb C}_2^2 
+ \left ( \frac{16525}{2 \cdot 3^6 \cdot (107)^4} -
\frac{58}{3^4 \cdot (107)^3}\zeta_3 \right ) {\mathbb C}_4 \cr\cr
&+& \left ( \frac{5}{3^7 \cdot (107)^3} 
-\frac{8}{3^6 \cdot (107)^3} \zeta_3 \right ) {\mathbb C}_2^3 
+ \left ( -\frac{37}{2 \cdot 3^6 \cdot (107)^3} + 
\frac{25}{2 \cdot 3^5 \cdot (107)^3}\zeta_3 \right ) 
{\mathbb C}_2{\mathbb C}_4 \cr\cr
&+& \left ( -\frac{1}{3^3 \cdot (107)^3} + 
\frac{16}{3^5 \cdot (107)^3}\zeta_3 \right ){\mathbb C}_{444} \ , 
\label{kappa3_generalbaryon}
\eeqs
\end{widetext} 
where $\zeta_s = \sum_{n=1}^\infty n^{-s}$ is the Riemann zeta function. 
In Eqs. (\ref{kappa1_generalbaryon})-(\ref{kappa3_generalbaryon}) we have
indicated the simple factorizations of the denominators.  The numerators do
not, in general, have such simple factorizations. 

In floating-point format, to the indicated precision, 
\beq
\kappa_{B,1} = (2.076843 \times 10^{-3}){\mathbb C}_2 \ ,
\label{kappa1_generalbaryon_num}
\eeq
\beqs
\kappa_{B,2} &=& -(2.329170 \times 10^{-4}){\mathbb C}_0 
+(1.364679 \times 10^{-4}){\mathbb C}_2 \cr\cr
&+& (1.078319 \times 10^{-6}){\mathbb C}_2^2 
- (5.391597 \times 10^{-6}){\mathbb C}_4  \ , 
\cr\cr
&&
\label{kappa2_generalbaryon_num}
\eeqs
and
\beqs
\kappa_{B,3} &=& -(0.721139 \times 10^{-6}){\mathbb C}_0 
-(0.376693 \times 10^{-6}){\mathbb C}_2 \cr\cr
&+&(0.681918 \times 10^{-6}){\mathbb C}_2^2
-(0.616147 \times 10^{-6}){\mathbb C}_4 \cr\cr
&-&(0.890178 \times 10^{-8}){\mathbb C}_2^3 + (2.975975 \times 10^{-8})
{\mathbb C}_2 {\mathbb C}_4 \cr\cr
&+&(0.343749 \times 10^{-7}){\mathbb C}_{444} \ . 
\label{kappa3_generalbaryon_num}
\eeqs
%


\section{Scheme-Independent Series 
Expansions for Anomalous Dimensions of Specific Baryon Operators}
\label{gamma_spcific_baryon_section}

In this section we present results for coefficients in scheme-independent
series expansions for the anomalous dimensions of specific baryon operators. 
These analytic results are new here.
The anomalous dimension of the baryon operator ${\cal
  O}^{(j_1,j_2)}_{\pm}$ is denoted $\gamma_B^{(j_1,j_2),\pm}$.  We express the
scheme-independent series exansion for this anomalous dimension as
\beq
\gamma_B^{(j_1,j_2),\pm} = \sum_{n=1}^\infty \kappa^{(j_1,j_2),\pm}_n \, 
\Delta_f^n 
\label{gamma_spins_deltaseries}
\eeq
The truncation of this infinite series to maximal power (order) $\Delta_f^p$ is
denoted $\gamma^{(j_1,j_2),\pm}_{B,\Delta_f^p}$. We note that numerical results
for the $\Delta_f$ series expansions for two of the four specific operators,
namely, ${\cal O}^{(\frac{1}{2},0)}_{\pm}$, were given previously in
\cite{psbaryon}.  Since they were based on the results of \cite{gracey_baryon},
they should be multipled by a factor of 2 \cite{gbe}.

We calculate the following: 
\beqs
\kappa^{(\frac{1}{2},0),+}_1 &=& \frac{8}{3 \cdot (107)} \cr\cr
                             &=& 2.492212 \times 10^{-2} 
\label{kappa1_hzplus}
\eeqs
\beqs
\kappa^{(\frac{1}{2},0),+}_2 &=& \frac{38758}{3^3 \cdot (107)^3} \cr\cr
                       &=& 1.171780 \times 10^{-2} 
\label{kappa2_hzplus}
\eeqs
\beqs
\kappa^{(\frac{1}{2},0),+}_3 &=& \frac{314021069}{3^5 \cdot (107)^5}
- \frac{97792}{3^3 \cdot (107)^4}\zeta_3 \cr\cr
  &=& 5.892227 \times 10^{-5} 
\label{kappa3_hzplus}
\eeqs
\beqs
\kappa^{(\frac{1}{2},0),-}_1 &=& \frac{8}{3 \cdot (107)} \cr\cr
                             &=& 2.492212 \times 10^{-2} 
\label{kappa1_hzminus}
\eeqs
\beqs
\kappa^{(\frac{1}{2},0),-}_2 &=& \frac{18626}{3^2 \cdot (107)^3} \cr\cr 
           &=& 1.689374 \times 10^{-3} 
\label{kappa2_hzminus}
\eeqs
\beqs
\kappa^{(\frac{1}{2},0),-}_3 &=& \frac{40784885}{3^3 \cdot (107)^5}
- \frac{70400}{3^3 \cdot (107)^4}\zeta_3 \cr\cr
         &=& 0.837892 \times 10^{-4} 
\label{kappa3_hzminus}
\eeqs
\beqs
\kappa^{(\frac{3}{2},0),+}_1 &=& -\frac{8}{3 \cdot (107)} \cr\cr
                             &=& -(2.492212 \times 10^{-2}) 
\label{kappa1_32zplus}
\eeqs
\beqs
\kappa^{(\frac{3}{2},0),+}_2 &=& -\frac{69574}{3^3 \cdot (107)^3} \cr\cr 
           &=& -(2.103448 \times 10^{-3}) 
\label{kappa2_32zplus}
\eeqs
\beqs
\kappa^{(\frac{3}{2},0),+}_3 &=& -\frac{32245429}{3^3 \cdot (107)^5}
+ \frac{1169920}{3^4 \cdot (107)^4}\zeta_3 \cr\cr
         &=& 4.730261 \times 10^{-5} 
\label{kappa3_32zplus}
\eeqs
\beqs
\kappa^{(1,\frac{1}{2}),-}_1 &=& -\frac{8}{3^2 \cdot (107)} \cr\cr
                             &=& -(0.830737 \times 10^{-2}) 
\label{kappa1_1hminus}
\eeqs
\beqs
\kappa^{(1,\frac{1}{2}),-}_2 &=& -\frac{62726}{3^4 \cdot (107)^3} \cr\cr 
           &=& -(6.321370 \times 10^{-4}) 
\label{kappa2_1hminus}
\eeqs
\beqs
\kappa^{(1,\frac{1}{2}),-}_3 &=& -\frac{314714429}{3^6 \cdot (107)^5}
+ \frac{178688}{3^3 \cdot (107)^4}\zeta_3 \cr\cr
         &=& 2.991050 \times 10^{-5} \ . 
\label{kappa3_1hminus}
\eeqs
As is evident from these results, all of the scheme-independent coefficients
$\kappa^{(\frac{1}{2},0),+}_n$ and $\kappa^{(\frac{1}{2},0),-}_n$ that have
been calculated, namely those for $n=1, \ 2, \ 3$, are positive. In contrast,
we find mixed signs for the scheme-independent coefficients
$\kappa^{(\frac{3}{2},0),+}_n$; while $\kappa^{(\frac{3}{2},0),+}_1$ and
$\kappa^{(\frac{3}{2},0),+}_2$ are negative, $\kappa^{(\frac{3}{2},0),+}_3$ is
positive, and similarly with the $\kappa^{(1,\frac{1}{2}),-}_n$ for $n=1, \ 2,
\ 3$.

\begin{figure}
  \begin{center}
    \includegraphics[height=6cm,width=8cm]{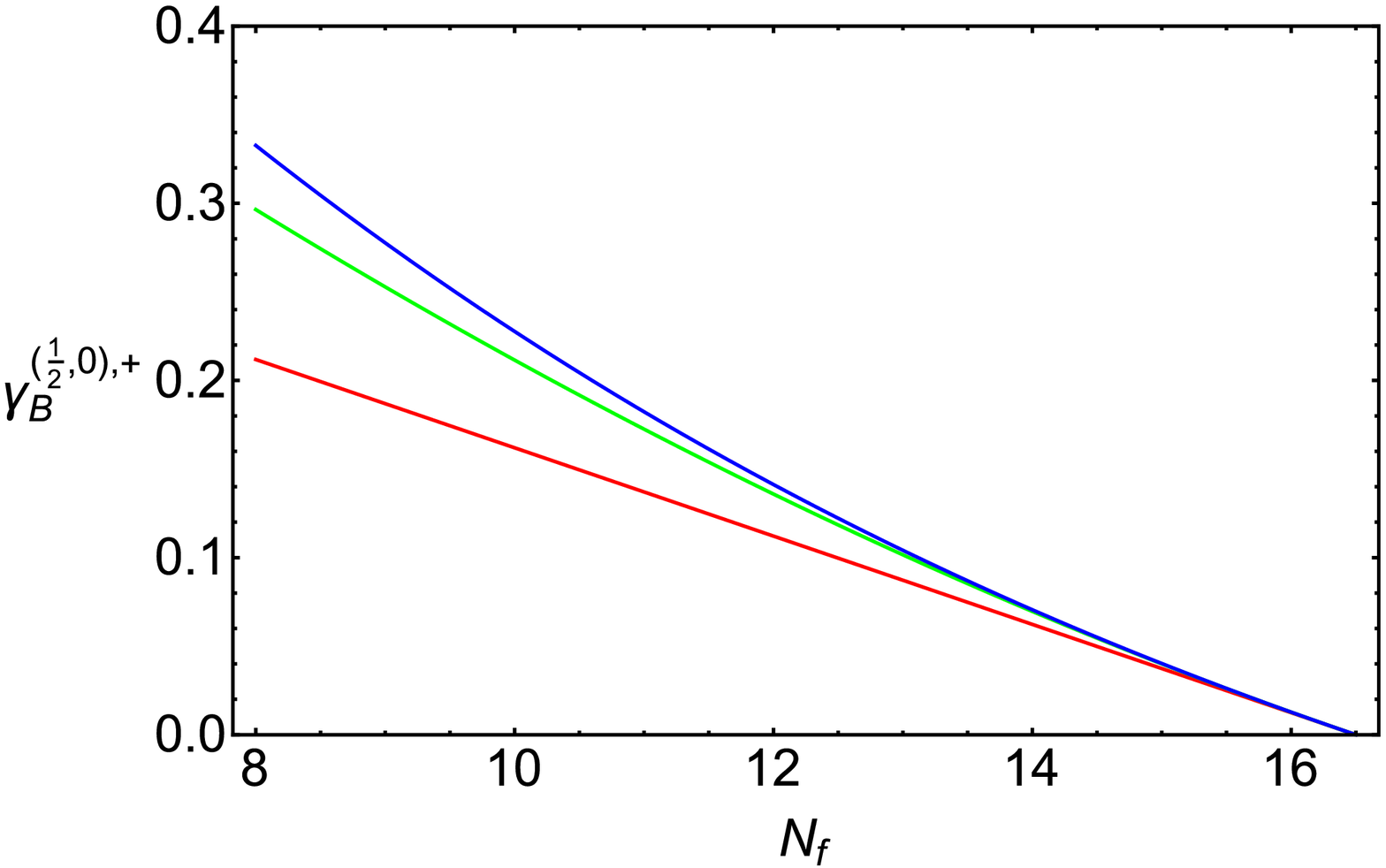}
  \end{center}
\caption{Plot of $\gamma_{B}^{(\frac{1}{2},0),+}$, 
as calculated in the scheme-independent
  series expansion to $O(\Delta_f^p)$ with $1 \le p \le 3$, as a function of 
$N_f$. The curves refer to the calculation to (a) $O(\Delta_f)$ (red)
$O(\Delta_f^2)$ (green), and $O(\Delta_f^3)$ (blue), with colors online.}
\label{gamma_B120p_plot}
\end{figure}

\begin{figure}
  \begin{center}
    \includegraphics[height=6cm,width=8cm]{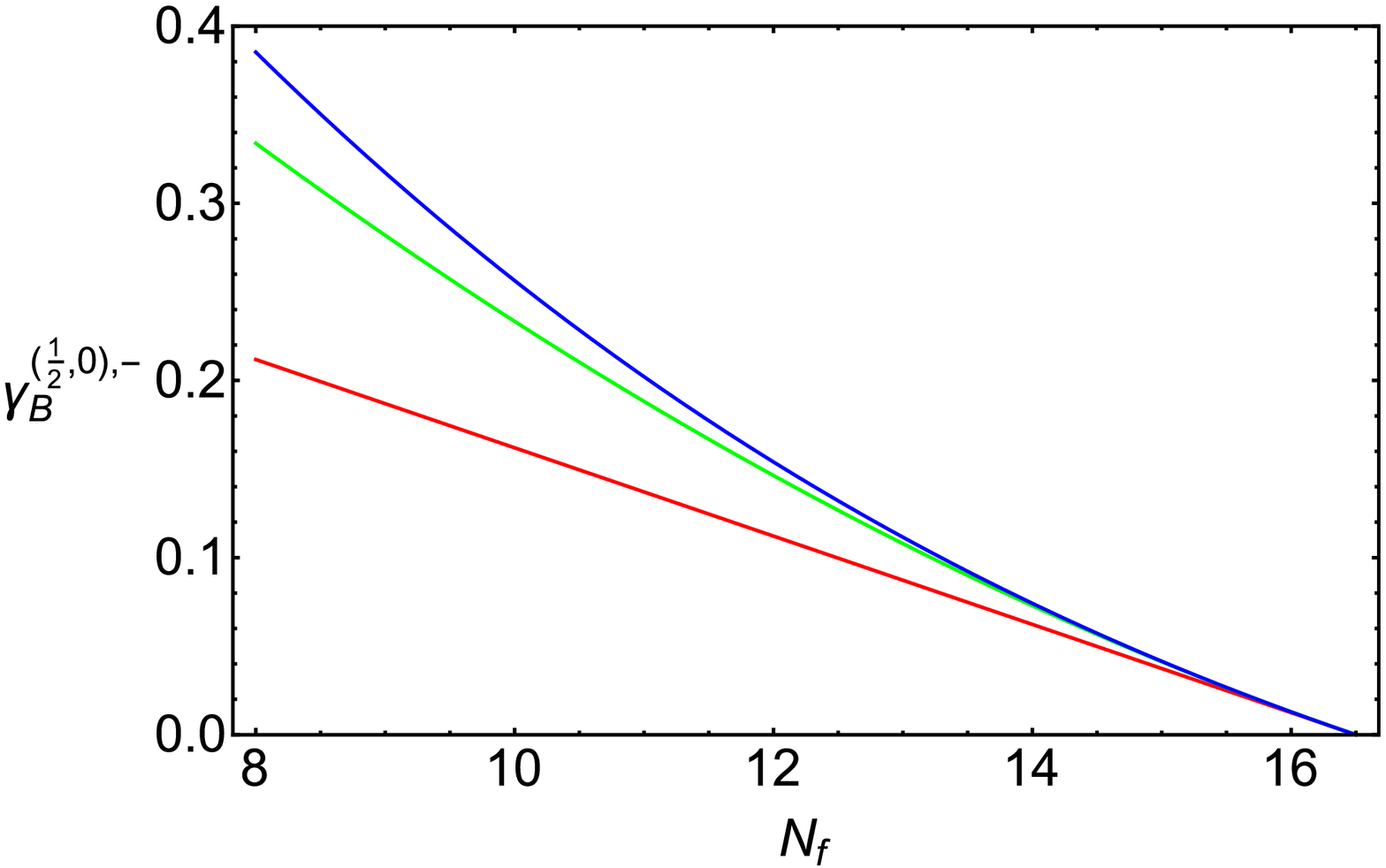}
  \end{center}
\caption{Plot of $\gamma_{B}^{(\frac{1}{2},0),-}$, 
as calculated in the scheme-independent
  series expansion to $O(\Delta_f^p)$ with $1 \le p \le 3$, as a function of 
$N_f$. The curves refer to the calculation to (a) $O(\Delta_f)$ (red)
$O(\Delta_f^2)$ (green), and $O(\Delta_f^3)$ (blue), with colors online.}
\label{gamma_B120m_plot}
\end{figure}

\begin{figure}
  \begin{center}
    \includegraphics[height=6cm,width=8cm]{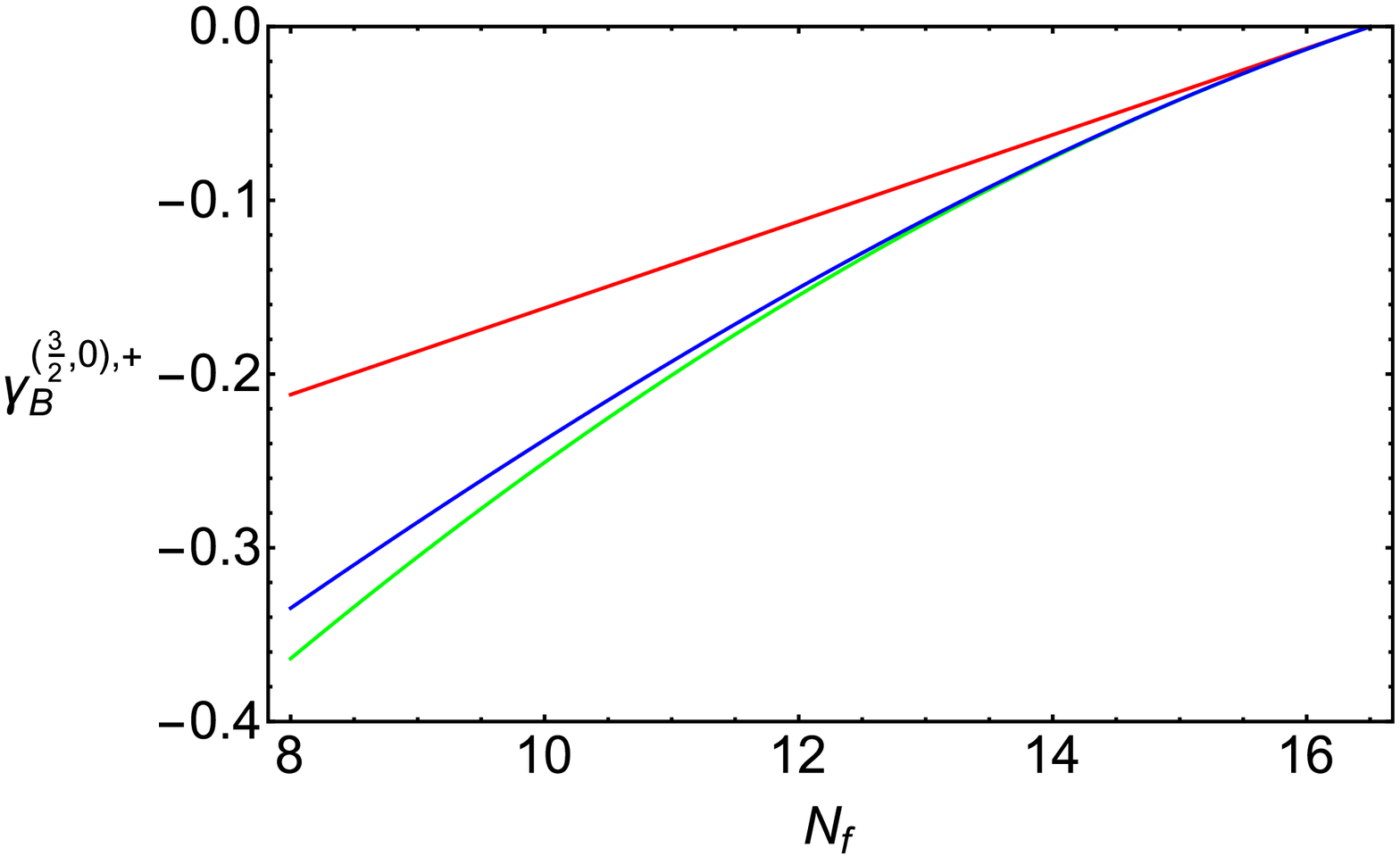}
  \end{center}
\caption{Plot of $\gamma_{B}^{(\frac{3}{2},0),+}$, 
as calculated in the scheme-independent
  series expansion to $O(\Delta_f^p)$ with $1 \le p \le 3$, as a function of 
$N_f$. The curves refer to the calculation to (a) $O(\Delta_f)$ (red)
$O(\Delta_f^2)$ (green), and $O(\Delta_f^3)$ (blue), with colors online.}
\label{gamma_B320p_plot}
\end{figure}

\begin{figure}
  \begin{center}
    \includegraphics[height=6cm,width=8cm]{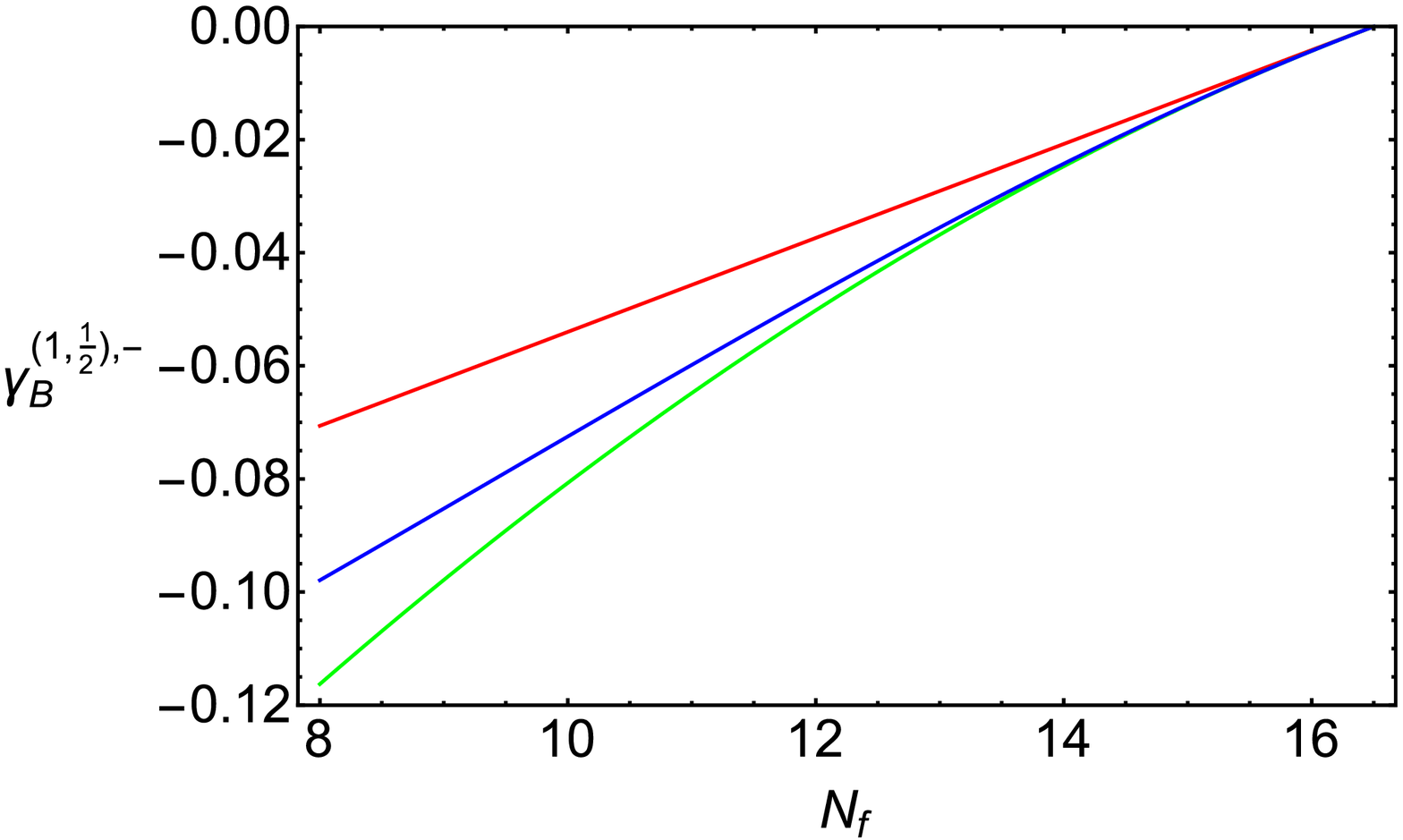}
  \end{center}
\caption{Plot of $\gamma_{B}^{(1,\frac{1}{2}),-}$, 
as calculated in the scheme-independent
  series expansion to $O(\Delta_f^p)$ with $1 \le p \le 3$, as a function of 
$N_f$. The curves refer to the calculation to (a) $O(\Delta_f)$ (red)
$O(\Delta_f^2)$ (green), and $O(\Delta_f^3)$ (blue), with colors online.}
\label{gamma_B112m_plot}
\end{figure}

In Figs. \ref{gamma_B120p_plot}-\ref{gamma_B112m_plot} we show curves of these
anomalous dimensions, and in Tables
\ref{gamma_B120p_table}-\ref{gamma_B112m_table} we list values of these
anomalous dimensions, as calculated to the various orders in $\Delta_f$ in our
scheme-independent expansions.

We comment further on the results for the coefficients 
$\kappa^{(\frac{1}{2},0),+}_n$ and $\kappa^{(\frac{1}{2},0),-}_n$ in the
respective scheme-independent series expansions for
$\gamma_B^{(\frac{1}{2},0),\pm}$. 
It will be recalled that an important property of the scheme-independent 
calculations of $\gamma_{\bar\psi\psi,IR}$ in
\cite{gtr,gsi,dex,dexs,dexl,dexo,dexss} is that (a) the 
coefficients $\kappa_1$ and $\kappa_2$ are manifestly positive, and (b) for all
groups and representations considered, $\kappa_3$ and $\kappa_4$ were also
found to be positive.  This result implied several monotonicity properties,
namely that (i) for a fixed truncation order $p$, the scheme-independent 
series expansion for $\gamma_{\bar\psi\psi,IR}$ is a monotonically increasing
function of $\Delta_f$, i.e., it increases monotonically with decreasing $N_f$,
and (ii) for a fixed value of $N_f$, the series calculation to $O(\Delta_f^p)$
is a monotonically increasing function of $p$. Indeed, as was noted in several
of these works, and was studied in detail in \cite{dexss}, the coefficients in
the corresponding scheme-independent expansions of anomalous dimensions of
composite holomorphic products of chiral superfields in ${\cal N}=1$
supersymmetric gauge theories are all positive.  

In view of these previous positivity findings, it is of considerable interest
that all of the $\kappa^{(\frac{1}{2},0),+}_n$ and 
$\kappa^{(\frac{1}{2},0),-}_n$ that have been calculated, namely those
for $j=1, \ 2, \ 3$, are positive, so the corresponding monotonicity
results apply for $\gamma_B^{(\frac{1}{2},0),\pm}$. 
These calculations to finite order in $O(\Delta_f)$
are expected to be most accurate for small $\Delta_f$, i.e., for $N_f$ slightly
below $N_u=16.5$, while higher-order corrections become progressively larger as
$N_f$ decreases toward the lower end of the NACP.  In \cite{gsi,dex,dexs,dexl}
these scheme-independent calculations were used to derive estimates of the
value of $N_f$ at the lower end of the NACP.  The method was to use the
unitarity lower bound $D_{\cal O} \ge 1$ for a Lorentz-scalar operator ${\cal
O}$ in a conformal field theory \cite{gammabound}.  From the basic definition
(\ref{dimop}), taking into account that the free-field (classical) dimension of
$\bar\psi\psi$ is $D_{\bar\psi\psi,cl.}=3$, there follows the upper bound
$\gamma_{\bar\psi\psi,IR} \le 2$.  Combining this with the above-mentioned
monotonicity results for the scheme-independent calculation of
$\gamma_{\bar\psi\psi,IR}$ yielded the estimate \cite{gsi,dex,dexs,dexl} that
the conformal non-Abelian Coulomb phase extends from $N_u=16.5$ down to
slightly above $N_f=8$, so the maximal value of $\Delta_f$ in this NACP, is
$(\Delta_f)_{max} \simeq 8$.

As was done for $\gamma_{\bar\psi\psi,IR}$ and $\beta'_{IR}$ in previous works
\cite{gtr,gsi,dex,dexl}, we may estimate the accuracy of these $O(\Delta_f^3)$
series calculations of $\gamma_B^{(\frac{1}{2},0),+}$ and
$\gamma_B^{(\frac{1}{2},0),-}$ in several ways. The first is to plot the
various truncations to $O(\Delta_f^p)$ with $p=1, \ 2, \ 3$ as functions of
$\Delta_f$, or equivalently, $N_f$ in the conformal regime (non-Abelian Coulomb
phase) and ascertain how close the curves are to each other. As expected, the
curves of $\gamma_B^{(\frac{1}{2},0),+}$, calculated to the higher two orders,
$O(\Delta_f^2)$ and $O(\Delta_f^3)$, remain close to each other over a larger
range, extending to lower $N_f$, than the corresponding curves 
calculated to the lower two orders, $O(\Delta_f)$ and $O(\Delta_f^2)$.  A
similar comment applies to the corresponding curves of 
$\gamma_B^{(\frac{1}{2},0),-}$ 

We recall that if a function $f(z)$ is analytic at $z=0$ and thus has a Taylor
series $f(z) = \sum_{n=1}^\infty s_n z^n$, then the ratio test states that the
series converges to the function $f(z)$ if $|z| < z_0$, where
\beq
z_0 = \lim_{n \to \infty} \frac{|s_n|}{|s_{n+1}|} \ .
\label{z0}
\eeq
Of course, even if these series expansions in powers of
$\Delta_f$ were Taylor series, it would not be possible to actually calculate
the limit (\ref{z0}), since we have only the first few coefficients.
Furthermore, the $\Delta_f$ expansion is not generically expected to be a
Taylor series, because the properties of the theory change qualitatively as
$N_f$ increases through $N_u$ and the theory becomes IR-free instead of
UV-free.  Nevertheless, a calculation of the first few ratios can give a rough
idea of the accuracy of a truncation of the series to a given order.
Accordingly, this was carried out for $\gamma_{\bar\psi\psi,IR}$ and
$\beta'_{IR}$ in \cite{gtr,gsi,dex,dexs,dexl,dexo}.  It was found that the
series expansions for $\gamma_{\bar\psi\psi,IR}$ to $O(\Delta_f^4)$ and
$\beta'_{IR}$ to $O(\Delta_f^5)$ were reasonably accurate over a substantial
portion of the NACP.

It is thus worthwhile to carry out the analogous calculation of ratios here for
$\gamma_B^{(\frac{1}{2},0),\pm}$. We find 
\beq
\frac{\kappa^{(\frac{1}{2},0),+}_1}
     {\kappa^{(\frac{1}{2},0),+}_2} = 21.27
\label{r12_plus}
\eeq
\beq
\frac{\kappa^{(\frac{1}{2},0),+}_2}
     {\kappa^{(\frac{1}{2},0),+}_3} = 19.89
\label{r23_plus}
\eeq
\beq
\frac{\kappa^{(\frac{1}{2},0),-}_1}
     {\kappa^{(\frac{1}{2},0),-}_2} = 14.75
\label{r12_minus}
\eeq
and
\beq
\frac{\kappa^{(\frac{1}{2},0),-}_2}
     {\kappa^{(\frac{1}{2},0),-}_3} = 20.16 \ . 
\label{r23_minus}
\eeq
These ratios are all substantially larger than $(\Delta_f)_{\rm max} \simeq 8$,
indicating that the scheme-independent series expansions for
$\gamma_B^{(\frac{1}{2},0),\pm}$ to $O(\Delta_f^3)$ may be reasonably accurate
over a substantial part of the NACP for this SU(3) theory.


\section{Unitarity Bounds on Anomalous Dimensions of Baryonic Operators} 
\label{gamma_bounds}

Since our scheme-independent series expansions for baryon operators apply at an
infrared fixed point in the non-Abelian Coulomb phase, where the theory is
conformally invariant, it is of interest to study how the resultant anomalous
dimensions compare with the unitarity bounds on a conformal field theory.  In
general \cite{gammabound}, for an operator ${\cal O}$ characterized by 
Lorentz spins $(j_1,j_2)$, unitarity in a conformal field theory requires that
the full scaling dimension $D_{\cal O}$ is bounded below according to 
\beq
D_{\cal O} \ge j_1 + j_2 + 1 \ . 
\label{dbound}
\eeq
For our case of SU(3), the free-field dimension of a baryon operator is
$D_{B,{\rm free}}=3(3/2) = 9/2$, so, with Eq. (\ref{dimop}), the lower bound
(\ref{dbound}) implies the upper bound on the anomalous dimension 
\beq
{\rm SU}(3): \quad \gamma_B^{(j_1,j_2)} \le \frac{7}{2} - (j_1+j_2) \ . 
\label{gammaboundb}
\eeq
Specifically, for the various operators considered here (suppressing $\pm$), 
\beq
\gamma_B^{(\frac{1}{2},0)} \le 3
\label{gammabound120}
\eeq
\beq
\gamma_B^{(\frac{3}{2},0)} \le 2
\label{gammabound320}
\eeq
and
\beq
\gamma_B^{(1,\frac{1}{2})} \le 2 \ . 
\label{gammabound112}
\eeq
For the present theory with gauge group SU(3) and $N_f$ fermions in the
fundamental representation, the previous work in \cite{gsi,dex,dexs,dexl} led
to the inference that the lower end of the NACP occurs at $N_{f,cr}$ around
8-9. In Fig. \ref{gamma_B120p_plot} and Fig. \ref{gamma_B120m_plot}, one can
see that our scheme-independent calculations of $\gamma_B^{(\frac{1}{2},0)+}$
and $\gamma_B^{(\frac{1}{2},0)-}$ to $O(\Delta_f^3)$ are well below the upper
bound of 3 in (\ref{gammabound120}).  Our results for
$\gamma_B^{(\frac{3}{2},0)+}$ and $\gamma_B^{(1,\frac{1}{2})-}$ are negative,
so they obviously also satisfy the respective upper bounds
(\ref{gammabound320}) and (\ref{gammabound112}).

The fact that these baryon anomalous dimensions, as calculated to
$O(\Delta_f^3)$, do not saturate their respective unitarity upper bounds as
$N_f$ decreases toward the lower end of the non-Abelian Coulomb phase is
reminiscent of the situation for an ${\cal N}=1$ supersymmetric gauge theory
with gauge group SU($N_c$) and $N_f$ pairs of chiral superfields, transforming
respectively as the representations $R$ and $\bar R$ of SU($N_c$), as studied
in \cite{dexss}. For this supersymmetric gauge theory, the only composite
chiral superfield for which the anomalous dimension saturates its unitarity
upper bound from conformal invariance as $N_f$ approaches the lower end of the
NACP from above is the gauge-invariant quadratic chiral superfield, which
contains the $\bar\psi\psi$ component field product. In contrast, (aside from
the pseudoreal case of SU(2)), a baryonic chiral superfield does not saturate
its unitarity upper bound from conformal invariance at the lower end of the
NACP \cite{dexss}.


\section{Schemes for Illustrative Calculations} 
\label{schemes_section}

In this section we review some background and methods relevant for our
calculations illustrating the scheme independence of the $\Delta_f$ series
expansions for $\gamma_{\bar\psi\psi,IR}$ and $\beta'_{IR}$.  We consider
several schemes for regularization and renormalization. We first discuss these
schemes. Recall that a common expression that one obtains from loop integrals
performed in $d$-dimensional spacetime is
\beq
\frac{\Gamma(2-(d/2))}{(4\pi)^{d/2}} \frac{1}{A^{(d/2)-2}} \ , 
\label{integral}
\eeq
where $\Gamma(z)$ is the Euler gamma function, and $A$ is a
denominator depending on some external momenta.  Defining $\epsilon=4-d$ and
expanding about $\epsilon=0$, using 
the Taylor-Laurent expansion of $\Gamma(z)$ about a pole at $z=0$, 
\beq
\Gamma(z) = \frac{1}{z} - \gamma_E + O(z) \ , 
\label{gammapole}
\eeq
Eq. (\ref{integral}) becomes 
\beq
\frac{1}{(4\pi)^2} \left [ \frac{2}{\epsilon} - \gamma_E + \ln(4\pi) - \ln A +
    O(\epsilon) \right ] \ , 
\label{integral_expansion}
\eeq
where
\beqs
\gamma_E &=& \lim_{n \to \infty} \left ( \sum_{k=1}^n \frac{1}{k}
- \ln n \right ) \cr\cr
         & \simeq & 0.5772157 
\label{gamma_euler}
\eeqs
In the minimal subtraction scheme MS \cite{ms}, one subtracts the pole term,
$2/\epsilon$.  In the modified minimal subtraction scheme $\overline{\rm MS}$ \
\cite{msbar}, one subtracts the pole term and also the two following terms,
namely the combination $2/\epsilon - \gamma_E + \ln(4\pi)$.  Both the MS and
$\overline{\rm MS}$ schemes are mass-independent and have the appeal that the
beta function and anomalous dimensions of gauge-invariant operators are
gauge-invariant.  As was noted above, the calculations of
\cite{gtr,gsi,dex,dexs,dexl,dexo,pgb,bpcgt} used this scheme, although the
resulting $\Delta_f$ expansions were proved to be scheme-independent.

In addition to the $\overline{\rm MS}$ scheme used in the previous work 
\cite{gsi,dex,dexs,dexl,dexo,pgb}, the
schemes that we use for our present illustrative demonstrations of scheme
independence of $\Delta_f$ expansions are 

\begin{enumerate}

\item The modified renormalization-invariant scheme (RI$^\prime$) 
\cite{ri_prime,gracey_ri_prime} 

\item The momentum subtraction scheme MOMggg defined by focusing on 
  the triple-gluon vertex \cite{celmaster,graceysimms_mom_omega}  

\item The momentum subtraction scheme MOMh defined by focusing on 
  the gluon-ghost-ghost vertex \cite{celmaster,graceysimms_mom_omega} 

\item The momentum subtraction scheme MOMq defined by focusing on the
  gluon-fermion-fermion vertex \cite{celmaster,graceysimms_mom_omega}
(indicated with the subscript $q$ for ``quark'')

\item The minimal momentum subtraction (mMOM) scheme \cite{mmom,gracey_mmom}.

\end{enumerate}

We write the conventional expansion of $\gamma_{\bar\psi\psi}$ as
\beq
\gamma_{\bar\psi\psi} = \sum_{\ell=1}^\infty c_\ell \, a^\ell \ . 
\label{gamma}
\eeq
where the $c_\ell$ are the $\ell$-loop coefficients and, where no confusion
will result, we set $c_\ell \equiv c_{\bar\psi\psi,\ell}$.  The one-loop
coefficient, $c_1=6C_f$, is scheme-independent, while the $c_\ell$ with $\ell
\ge 2$ are scheme-dependent \cite{gross75}. The evaluation of the $n$-loop
truncation of (\ref{gamma}) at the IRFP is obtained by substituting
$\alpha=\alpha_{IR,n\ell}$ and is denoted $\gamma_{IR,n\ell}$.

Concerning the beta function (\ref{beta}), the one-loop coefficient, $b_1$
\cite{b1}, is scheme-independent.  In mass-independent schemes, the two-loop
coefficient, $b_2$ \cite{b2}, is also independent of the specific scheme
\cite{gross75}.  We have mentioned above the calculations of $b_3$ \cite{b3},
$b_4$ \cite{b4}, and $b_5$ \cite{b5su3,b5} in the $\overline{\rm MS}$ scheme.
As noted, the $c_\ell$ were calculated to four-loop order \cite{c4} and to
five-loop order in \cite{c5}, in the $\overline{\rm MS}$ scheme \cite{c5n}.

The $b_\ell$ and $c_\ell$ have been calculated to four-loop order in the
RI$^\prime$ scheme \cite{gracey_ri_prime} and the minimial MOM (mMOM)
scheme \cite{gracey_mmom}. Additional calculations in generalized MOM schemes
were presented in \cite{graceysimms_mom_omega}. 
A comparison of conventional calculations of
  $\alpha_{IR,n\ell}$ and $\gamma_{IR,n\ell}$ was given up to the four-loop
  order in \cite{tr_rip},\cite{tr_mmom}, \cite{tr_ripmom}, and
  \cite{bzmom2015}.  An important aspect in which the RI$^\prime$ and MOM 
schemes differ with the $\overline{\rm MS}$ scheme is that beyond the lowest
orders, the $b_\ell$ and $c_\ell$ are gauge-dependent.  We consider a covariant
gauge-fixing term so that the gauge part of the Lagrangian is 
(with our $(+---)$ metric) 
\beq
{\cal L}_{\rm gauge} = -\frac{1}{4}F^a_{\mu\nu}F^{\mu\nu,a} - 
\frac{1}{2\xi}(\partial^\mu A^a_\mu)^2 + {\rm F.P.}, 
\label{lgauge}
\eeq
where
\beq
F^a_{\mu\nu} = \partial_\mu A^a_\nu - \partial_\nu A^a_\mu + 
gf^{abc}A_\mu^b A_\nu^c
\label{fmunu}
\eeq
is the field-strength tensor, with $a=1,..,o(G)$ is the group index, $o(G)$
is the order of the gauge group, $f^{abc}$ are the structure constants of the
Lie algebra of $G$, and F.P. denote Faddeev-Popov terms.  The
gauge field propagator is thus 
\beq
\Delta^{ab}_{\mu\nu}(k) = -\frac{ \delta^{ab} \, \left [ g_{\mu\nu} -
 (1-\xi)\frac{k_\mu k_\nu}{k^2} \right ]}{k^2} \ . 
\label{gaugepropagator}
\eeq
The Landau gauge corresponds to $\xi=0$, where this propagator is transverse,
i.e., $k^\mu \Delta^{ab}_{\mu\nu}(k)=0$.  In these other schemes, the gauge
parameter $\xi$ also depends on the Euclidean scale $\mu$, and so there is an
associated function that measures this dependence, namely
\beq
\beta_\xi = \frac{d\xi}{d\ln\mu} \ . 
\label{betaxidef}
\eeq
We write the series expansion for this in powers of the coupling as
\beq
\beta_\xi = - 2\xi \sum_{\ell=1}^\infty b_{\xi,\ell} \, a^\ell \ . 
\label{betaxi}
\eeq
Evidently, the situation is the simplest in Landau gauge, since in this gauge,
$\beta_\xi=0$ and the gauge parameter is independent of the Euclidean scale.
The value of $\alpha$ at the IR zero of $\beta_\alpha$ and the resultant value
of $\gamma_{\bar\psi\psi,IR}$ were calculated in Landau gauge at the
three-loop level in the RI$^\prime$ scheme in \cite{tr_rip} and in the minimal
MOM (mMOM) scheme in \cite{tr_mmom}, and at the four-loop level in
\cite{tr_ripmom}.  We recall the procedure for this calculation.  One looks for
a physically acceptable simultaneous solution to the two coupled equations
\beq
\beta_\alpha(\alpha,\xi)=0, \quad\quad \beta_\xi(\alpha,\xi)=0 \ , 
\label{coupledequations}
\eeq
where we have explicitly indicated the dependence of $\beta_\alpha$ and
$\beta_\xi$ on the variables $\alpha$ and $\xi$.  Because $\beta_\xi$ is
proportional to $\xi$, one is always guaranteed to find a solution with
$\xi=0$.  That is, if $\xi=0$ at some value $\mu=\mu_0$, then $\xi=0$ for all
$\mu$.  This was the basis for the choice of Landau gauge in
Refs. \cite{tr_rip}, \cite{tr_mmom}, and \cite{tr_ripmom}.  As was discussed in
\cite{tr_ripmom}, there also exist fixed points for which $\xi \ne 0$, but
these solutions are on a different footing from the $\xi=0$ solution. As was
noted in \cite{tr_rip}, at the two-loop level in the mMOM scheme, there is also
an IRFP with $\xi_{2\ell}=-3$, and calculations at the three-loop level exhibit
an IRFP with $\xi_{3\ell}$ near to this value (see also \cite{alm}). A list of
the $b_\ell$, $b_{\xi,\ell}$, and $c_\ell$ for general $\xi$, with $1 \le \ell
\le 3$ in the mMOM scheme was given in \cite{tr_rip} and a list of the
$b_\ell$, $b_{\xi,\ell}$, and $c_\ell$ for $\xi=0$, i.e., Landau gauge, with $1
\le \ell \le 4$ was given in \cite{tr_ripmom} for the RI$^\prime$ and mMOM
schemes. We will also remark on the general case in which $\xi$ is not
necessarily zero. The corresponding expressions for the $b_\ell$, $c_\ell$, and
$b_{\xi,\ell}$ are too long and complicated to include here; they have been
given, for example, as external files with the arXiv version of
\cite{graceysimms_mom_omega}.  An important difference between the $c_\ell$ in
the RI$^\prime$ scheme and the $b_\ell$ and $c_\ell$ in the MOM schemes, as
contrasted with the $b_\ell$ and $c_\ell$ in the $\overline{\rm MS}$ scheme is
that in the non-MS schemes, these coefficients depend on a number of additional
mathematical functions and constants.  For example, as was discussed in
\cite{bzmom2015}, at the four-loop level, in addition to the dependence on the
group invariants $C_A$, $C_f$, and $T_f$, the $b_\ell$ and $c_\ell$ in the
$\overline{\rm MS}$, RI$^\prime$, and mMOM schemes contain dependence on the
quantities
\beq
\{ {\mathbb Q}, \ \zeta_3, \ \zeta_5 \} \ . 
\label{msset}
\eeq
Note that $\zeta_m$ with even $m=2r$ are proportional to $\pi^{2r}$:
\beq
\zeta_{2r}=\frac{(-1)^{r+1}B_{2r}(2\pi)^{2r}}{2(2r)!} \ ,
\label{zetaeven}
\eeq
where the $B_n$ are the Bernoulli numbers, defined by
\beq
\frac{t}{e^t-1} = \sum_{n=0}^\infty B_n \frac{t^n}{n!} \ ,
\label{bernoulli}
\eeq
so listing $\pi^2$ in (\ref{msset}) is equivalent to listing $\zeta_2$, etc.
In contrast, even at the lower, three-loop level, $b_\ell$ and $c_\ell$ in the
other MOM schemes have a considerably more complicated form, since they depend
on the following set of mathematical functions and constants:
\begin{widetext}
\beq
\{ {\mathbb Q}, \ \pi^2, \ \zeta_3, \ \pi^4, \ \psi'(1/3), \ \psi'''(1/3), \ 
s_2(\pi/k), \ s_3(\pi/k), \ \frac{\pi \ln(3)}{\sqrt{3}}, \ 
\frac{\pi \ln(3)^2}{\sqrt{3}}, \ \frac{\pi^3}{\sqrt{3}} \} \ , 
\label{genmomset}
\eeq
\end{widetext}
where here $k$ takes the values $k=2$ and $k=6$; $\psi(s)$ is the Euler $\psi$
function, $\psi(s) = d\ln[\Gamma(s)]/ds$, 
$\psi'(s) = d\psi(s)/ds$, and $s_n(z)$ is defined as
\beq
s_n(z) = \frac{1}{\sqrt{3}} \, {\rm Im} \left [ {\rm Li}_n \left ( 
\frac{e^{iz}}{\sqrt{3}} \right ) \right ] \ , 
\label{sn}
\eeq
where ${\rm Li}_n(z)$ is the polylogarithm function,
\beq
{\rm Li}_n(z) = \int_0^z \, \frac{{\rm Li}_{n-1}(z)}{t} \, dt
\label{lin}
\eeq
with ${\rm Li}_0(z)=z/(1-z)$ and ${\rm Li}_1(z)=-\ln(1-z)$.  For $|z| \le 1$,
this function has the series representation
\beq
{\rm Li}_n(z) = \sum_{j=1}^\infty \frac{z^j}{j^n} \ , n=2, \ 3,...
\label{linseries}
\eeq

The calculation of the coefficient $d_n$ in Eq. (\ref{betaprime_deltaseries})
requires, as input, the $\ell$-loop coefficients $b_\ell$ with $1 \le \ell \le
n$.  The calculation of the coefficient $\kappa_n$ in
Eq. (\ref{gamma_deltaseries}) requires, as inputs, the values of the $b_\ell$
for $1 \le \ell \le n+1$, and the $\ell$-loop coefficients $c_\ell$ in
Eq. (\ref{gamma}) with $1 \le \ell \le n$.  

In addition to our explicit demonstration that different schemes yield the same
values for the coefficients $d_n$ and $\kappa_n$ in the scheme-independent
expansions (\ref{betaprime_deltaseries}) and (\ref{gamma_deltaseries}), our
work shows that the full physical content of these scheme-independent
coefficients is derived from the use of the simplest scheme, namely
$\overline{\rm MS}$.  Thus, there is a huge cancellation of the additional
mathematical functions and quantities in (\ref{genmomset}) in the
scheme-independent coefficients $d_n$ and $\kappa_n$.  On the one hand, one may
take the view that this had to be true, since a rigorous proof was given
already that these coefficients are scheme-independent and their values were
therefore already completely determined from the calculations in
\cite{gtr}-\cite{dexl} in the $\overline{\rm MS}$ scheme.  But nevertheless,
our explicit demonstration of the cancellation is quite a striking result.


\section{Scheme-Independent Expansion of $\gamma_{\bar\psi\psi,IR}$}
\label{kappa_section}

The coefficients $\kappa_n$ in the scheme-independent expansion of 
$\gamma_{\bar\psi\psi,IR}$ in powers of $\Delta_f$, Eq. 
(\ref{gamma_deltaseries}), were calculated for a gauge group $G$ with
$N_f$ fermions in a representation $R$ up to $n=3$ in \cite{gtr} and up to
$n=4$ in \cite{dexs,dexl}.  (The coefficient $\kappa_4$ was calculated for 
$G={\rm SU}(3)$ and $R=F$ in \cite{gsi}.)  For example, the first two of these
coefficients are
\beq
\kappa_1 = \frac{8C_fT_f}{C_A(7C_A+11C_f)}
\label{kappa1}
\eeq
and
\beq
\kappa_2 = \frac{4C_fT_f^2(5C_A+88C_f)(7C_A+4C_f)}{3C_A^2(7C_A+11C_f)^3} \ . 
\label{kappa2}
\eeq
For the present work we have explicitly verified that we obtain the same
results for these $\kappa_n$ using the RI$^\prime$, mMOM, and other MOM
schemes. We have carried out this check to the highest order possible with
existing inputs available in these schemes, i.e., to order $n=3$. 


\section{Scheme-Independent Expansion of $\beta'_{IR}$}
\label{betaprime_section}

The derivative $\beta'_{IR}$ is an important physical quantity characterizing
the conformal field theory at $\alpha_{IR}$. For general gauge group $G$ with
$N_f$ fermions in a general representation $R$, the scheme-independent 
coefficients $d_n$ were
calculated up to $n=4$ in \cite{dex} and up to $n=5$ in \cite{dexs,dexl}. 
The first two nonzero coefficients are 
\beq
d_2 = \frac{2^5 T_f^2}{3^2 C_A(7C_A+11C_f) } 
\label{d2}
\eeq
and
\beq
d_3 = \frac{2^7 T_f^3(5C_A+3C_f)}{3^3 C_A^2 (7C_A+11C_f)^2} \ . 
\label{d3}
\eeq
We have explicitly verified that we obtain the same results for $d_n$ with the
RI$^\prime$, mMOM, and other MOM schemes. We have carried out this check to 
the highest order possible with existing inputs available in these 
schemes, i.e., to order $n=4$. 


\section{Conclusions}
\label{conclusion_section}

In conclusion, in this paper we have presented the first analytic
scheme-independent expansions to $O(\Delta_f^3)$ for the anomalous dimensions
of a variety of (gauge-invariant) baryon operators at an infrared fixed point
of an asymptotically free SU(3) gauge theory with $N_f$ fermions in the
fundamental (triplet) representation.  Furthermore, for an asymptotically free
theory with a general gauge group $G$ and $N_f$ fermions in a general
representation $R$ of $G$, we have given explicit illustrative demonstrations
of the scheme independence of $\gamma_{\bar\psi\psi,IR}$ and $\beta'_{IR}$ at
an IRFP.  Although this scheme independence had been proved rigorously earlier,
it is worthwhile to see how different schemes yield identical results for the
coefficients in the scheme-independent expansions.  We have carried out these
calculations for the RI$^\prime$ and several MOM schemes.


\begin{acknowledgments}

  We thank R.M. Simms and L. Vecchi for useful discussions. This research was
  supported in part by U.K.  Science and Technology Facilities Council (STFC)
  through the Consolidated Grant ST/L000431/1 and a Deutsche
  Forschungsgemeinschaft (DFG) Mercator Fellowship (J.A.G.), the Danish
  National Research Foundation grant DNRF90 to CP$^3$-Origins at SDU (T.A.R.)
  and the U.S. National Science Foundation Grant NSF-PHY-16-1620628 (R.S.).
  This research was initiated at the Simons Workshop \cite{simons} in Jan.,
  2018, and the authors are grateful to the Simons Foundation for funding this
  workshop.

\end{acknowledgments}



\begin{table}
  \caption{\footnotesize{Values of 
$\gamma_{B,\Delta^p}^{\left( \frac{1}{2},0 \right),+}$ with 
$1\leq p \leq 3$.  }}
\begin{center}
\begin{tabular}{|c|c|c|c|} \hline\hline
$N_f$ & 
$\gamma_{B,\Delta^1}^{\left( \frac{1}{2},0 \right),+}$ & 
$\gamma_{B,\Delta^2}^{\left( \frac{1}{2},0 \right),+}$ & 
$\gamma_{B,\Delta^3}^{\left( \frac{1}{2},0 \right),+}$ \\
 \hline\hline
$8$  & $0.212$ & $0.296$ & $0.333$   \\
\hline
$9$   & $0.187$  & $0.253$ & $0.278$  \\
\hline
$10$ & $0.162$ & $0.212$ & $0.228$   \\
 \hline
$11$ & $0.137$ & $0.173$ & $0.182$  \\
 \hline
$12$ & $0.112$ & $0.136$ & $0.141$  \\
 \hline
$13$ & $0.0872$ & $0.102$ & $0.104$  \\
 \hline
$14$ & $0.0623$ & $0.0696$ & $0.0705$  \\
 \hline
$15$ & $0.0374$ & $0.0400$ & $0.0402$    \\
 \hline
$16$ & $0.0125$ & $0.0128$ & $0.0128$  \\
\hline\hline
\end{tabular}
\end{center}
\label{gamma_B120p_table}
\end{table}

\begin{table}
  \caption{\footnotesize{Values of 
$\gamma_{B,\Delta^p}^{\left( \frac{1}{2},0 \right),-}$ with 
$1\leq p \leq 3$. }}
\begin{center}
\begin{tabular}{|c|c|c|c|} \hline\hline
$N_f$ & 
$\gamma_{B,\Delta^1}^{\left( \frac{1}{2},0 \right),-}$ & 
$\gamma_{B,\Delta^2}^{\left( \frac{1}{2},0 \right),-}$ & 
$\gamma_{B,\Delta^3}^{\left( \frac{1}{2},0 \right),-}$ \\
 \hline\hline
$8$  & $0.212$ & $0.334$ & $0.385$ \\
\hline
$9$   & $0.187$ & $0.282$ & $0.317$   \\
\hline
$10$ & $0.162$ & $0.233$ & $0.256$  \\
 \hline
$11$ & $0.137$ & $0.188$ & $0.202$ \\
 \hline
$12$ & $0.112$ & $0.146$ & $0.154$  \\
 \hline
$13$ & $0.0872$ & $0.108$ & $0.112$ \\
 \hline
$14$ & $0.0623$ & $0.0729$ & $0.0742$ \\
 \hline
$15$ & $0.0374$ & $0.0412$ & $0.0415$  \\
 \hline
$16$ & $0.0125$ & $0.0129$ & $0.0129$  \\
\hline\hline
\end{tabular}
\end{center}
\label{gamma_B120m_table}
\end{table}

\begin{table}
  \caption{\footnotesize{Values of 
$\gamma_{B,\Delta^p}^{\left( \frac{3}{2},0 \right),+}$ with 
$1\leq p \leq 3$.  }}
\begin{center}
\begin{tabular}{|c|c|c|c|} \hline\hline
$N_f$ & 
$\gamma_{B,\Delta^1}^{\left( \frac{3}{2},0 \right),+}$ & 
$\gamma_{B,\Delta^2}^{\left( \frac{3}{2},0 \right),+}$ & 
$\gamma_{B,\Delta^3}^{\left( \frac{3}{2},0 \right),+}$ \\
 \hline\hline
$8$  & $-0.212$ & $-0.364$ & $-0.335$  \\
\hline
$9$   & $-0.187$ & $-0.305$ & $-0.285$  \\
\hline
$10$ & $-0.162$ & $-0.251$ & $-0.238$   \\
 \hline
$11$ & $-0.137$ & $-0.201$ & $-0.193$  \\
 \hline
$12$ & $-0.112$ & $-0.155$ & $-0.150$  \\
 \hline
$13$ & $-0.0872$ & $-0.113$ & $-0.111$  \\
 \hline
$14$ & $-0.0623$ & $-0.0755$ & $-0.0747$  \\
 \hline
$15$ & $-0.0374$ & $-0.0421$ & $-0.0420$  \\
 \hline
$16$ & $-0.0125$ & $-0.0130$ & $-0.0130$  \\
\hline\hline
\end{tabular}
\end{center}
\label{gamma_B320p_table}
\end{table}

\begin{table}
  \caption{\footnotesize{Values of 
$\gamma_{B,\Delta^p}^{\left( 1, \frac{1}{2} \right),-}$ with 
$1\leq p \leq 3$.  }}
\begin{center}
\begin{tabular}{|c|c|c|c|} \hline\hline
$N_f$ & 
$\gamma_{B,\Delta^1}^{\left( 1, \frac{1}{2} \right),-}$ & 
$\gamma_{B,\Delta^2}^{\left( 1, \frac{1}{2} \right),-}$ & 
$\gamma_{B,\Delta^3}^{\left( 1, \frac{1}{2} \right),-}$ \\
 \hline\hline
$8$  & $-0.0706$ & $-0.117$ & $-0.0979$  \\
 \hline
$9$   & $-0.0623$ & $-0.0979$ & $-0.0852$  \\
\hline
$10$ & $-0.0540$ & $-0.0807$ & $-0.0725$   \\
 \hline
$11$ & $-0.0457$ & $-0.0648$ & $-0.0598$   \\
 \hline
$12$ & $-0.0374$ & $-0.0502$ & $-0.0475$  \\
 \hline
$13$ & $-0.0291$ & $-0.0368$ & $-0.0355$  \\
 \hline
$14$ & $-0.0208$ & $-0.0247$ & $-0.0243$  \\
 \hline
$15$ & $-0.0125$ & $-0.0139$ & $-0.0138$  \\
 \hline
$16$ & $-0.00415$ & $-0.00431$ & $-0.00431$  \\
\hline\hline
\end{tabular}
\end{center}
\label{gamma_B112m_table}
\end{table}


\begin{thebibliography}{99}
\bibitem{fm}
Our assumption that the fermions are massless does not entail any 
loss of generality, since if a fermion had a nonzero mass $m_0$, it would be
integrated out of the low-energy effective field theory at scales $\mu <
m_0$, and hence would not affect the properties at the infrared fixed point.
\bibitem{casimir}
%
  For the representation $R$ of the Lie group $G$, the quadratic Casimir
  invariant $C_2(R)$ is given by $T^a_RT^a_R = C_2(R) I$, where $I$ is the
  ${\rm dim}(R) \times {\rm dim}(R)$ identity matrix.  For $R$ equal to the
  adjoint representation, we denote $C_2(adj) \equiv C_2(G) \equiv C_A$.  The
  trace invariant is ${\rm Tr}_R(T^a_R T^b_R) = T(R)\delta^{ab}$.  For a
  fermion $f$ transforming according to a representation $R$, we often use the
  equivalent compact notation $T_f \equiv T(R)$ and $C_f \equiv C_2(R)$.  We
  use standard normalizations, so that for SU($N_c$), $C_A=N_c$ and for $R$
  equal to the fundamental representation, $C_f=(N_c^2-1)/(2N_c)$ and
  $T_f=1/2$.
\bibitem{nfintegral}
%
We often implicitly consider $N_f$ to be generalized from the non-negative
integers to the non-negative real numbers. If an expression for $N_f$
formally evaluates to a non-integral real value, it is understood
implicitly that one infers an appropriate integral value from it.
\bibitem{gbe}
%
  With the definition (\ref{dimop}), our Eq. (\ref{gamma_generalbaryon})
  involves a multiplication by a factor of 2 relative to Eq. (6.13) in
  \cite{gracey_baryon}.
\bibitem{sksb}
%
In contrast, in the QCD-like phase with confinement, even if the two-loop beta
function has an IR zero, this is not exact.  The reason is that there is
spontaneous chiral symmetry breaking (S$\chi$SB) in this QCD-like phase, which
produces dynamical masses for the fermions. Consequently, the fermions are
integrated out of the low-energy effective theory as one evolves further toward
the infrared, and the beta function becomes that of a pure gauge theory, which
has no IR zero.
\bibitem{c5n}
%
  Parenthetically, we note that although the conventional calculation of the
  $\bar\psi\psi$ anomalous dimension to five-loop order in \cite{flir} used
  $c_5$ from \cite{c5}, the calculation of the scheme-independent coefficients
  $\kappa_n$ \cite{gtr,gsi,dexs,dexl,dexo,pgb}, which have been carried out up
  to $n=4$ order for general $G$ and $R$, did not make use of $c_5$. The reason
  is that, as mentioned in the text, the calculation of $\kappa_n$ requires, as
  input, the $b_\ell$ with $1 \le \ell \le n+1$ and the $c_\ell$ with $1 \le
  \ell \le n$.  Hence, in order to use $c_5$ to calculate $\kappa_5$, one would
  also need $b_6$, but the $b_\ell$ have only been calculated up to $\ell=5$
  loop order \cite{b5su3,b5}.
\end{thebibliography}
\end{document}